# Straight versus Spongy – Effect of Tortuosity on Polymer Imbibition into Nanoporous Matrices Assessed by Segmentation-Free Analysis of 3D Sample Reconstructions

Fernando Vazquez Luna,[#] Anjani K. Maurya,[%] Juliana Martins de Souza e Silva,[&,†] Guido Dittrich,[§] Theresa Paul,[$] Dirk Enke,[$] Patrick Huber,[§] Ralf Wehrspohn,[&] Martin Steinhart[*,#]

[#] Institut für Chemie neuer Materialien and CellNanOs, Universität Osnabrück, Barbarastr. 7, 49069 Osnabrück, Germany

[%] Stanford Synchrotron Radiation Lightsource, SLAC National Accelerator Laboratory, Menlo Park, CA, 94025 USA

[&] Martin-Luther-Universität Halle-Wittenberg, Institut für Physik, Fachgruppe Mikrostrukturbasiertes Materialdesign, Heinrich-Damerow-Str. 4, 06120 Halle, Germany

[§] Hamburg University of Technology, Institute for Materials and X-Ray Physics, 21073 Hamburg, Germany and Center for X-ray and Nano Science CXNS, Deutsches Elektronen-Synchrotron DESY, Germany and Hamburg University, Center for Hybrid Nanostructures CHyN, 22607 Hamburg, Germany

[$] Institute of Chemical Technology, Universität Leipzig, Linnéstraße 3, 04103 Leipzig, Germany

**Abstract**
We comparatively analyzed imbibition of polystyrene (PS) into two complementary pore models having pore diameters of ~380 nm and hydroxyl-terminated inorganic-oxidic pore walls, controlled porous glass (CPG) and self-ordered porous alumina (AAO), by X-ray computed tomography and EDX spectroscopy. CPG contains continuous spongy-tortuous pore systems. AAO containing arrays of isolated straight cylindrical pores is a reference pore model with a tortuosity close to 1. Comparative evaluation of the spatiotemporal imbibition front evolution yields important information on the pore morphology of a probed tortuous matrix like CPG and on the imbibition mechanism. To this end, pixel brightness dispersions in tomographic 3D reconstructions and 2D EDX maps of infiltrated AAO and CPG samples were condensed into 1D brightness dispersion profiles normal to the membrane surfaces. Their statistical analysis yielded positions and widths of the imbibition fronts without segmentation or determination of pore positions. The retardation of the imbibition front movement with respect to AAO reference samples may be used as a descriptor for the tortuosity of a tested porous matrix. The velocity of the imbibition front movements in CPG equaled two-thirds of the velocity of the imbibition front movements in AAO. Moreover, the dynamics of the imbibition front broadening discloses whether porous matrices are dominated by cylindrical neck-like pore segments or by nodes. Independent single-meniscus movements in cylindrical AAO pores result in faster imbibition front broadening than in CPG, in which a morphology dominated by nodes results in slower cooperative imbibition front movements involving several menisci. The results presented here may be relevant to applications including printing and adhesive bonding as well as to the optimization of production and properties of engineering, construction and hybrid materials.

**Introduction**

Imbibition of nanoporous media by invading fluids[1] is of significant relevance to a diverse variety of application fields ranging from geophysics[2] to carbon dioxide trapping[3] to inkjet printing[4] to dentistry.[5] Furthermore, imbibition underlies the preparation of functional nanostructured composite and hybrid materials,[6] including optical hybrids,[7,8] solar cells[9] and active ingredient depots.[10] The simplest imbibition model scenario is the invasion of straight cylindrical pores by fluids wetting the pore walls under conditions, where gravitation can be neglected. This scenario is commonly described by power laws relating the single-pore imbibition length $L_s$ (the length of the pore segment filled with the invading liquid) to the elapsed imbibition time $t_i$:

$$L_s = v \cdot t_i^n \quad \text{(Equation 1)}$$

The classical Lucas-Washburn model[11,12] assuming lamellar flow of an invading fluid through ideal cylindrical capillaries uniform in diameter predicts a value of 0.5 for the exponent *n* of *t$_i$*. This time scaling results from a dynamic balance of a constant driving Laplace pressure at the advancing imbibition front and a linear increase in viscous drag in the infiltrated part of the cylindrical capillary behind the imbibition front. Thus, any changes in the capillarity of the liquid, the liquid-solid interaction or the hydraulic permeability as a function of time during the imbibition process can lead to deviations from the classical Lucas-Washburn scenario. Experimentally observed deviations from the Lucas-Washburn model have also been traced to "dead layers" of adsorbed molecules of the invading species on the pore walls,[13-16] as well as to thermal fluctuations, van der Waals forces and hydrodynamic slippage.[17,18] Moreover, single-pore imbibition is characterized by a broad phenomenology of transient nanoscopic single-pore imbibition morphologies often characterized by the presence of precursor films.[19,20] Attempts to correspondingly adapt the Lucas-Washburn equation predominantly addressed the nature of the preexponential factor $v$.[21-23] On the other hand, departures from an ideal cylindrical pore geometry can lead to deviations from the classical proportionality of the average imbibition front position $L_a$ to $t_i^{1/2}$. Already conical pore shapes may result in an exponent *n* of *t$_i$* of 0.25.[24] Real-life porous model matrices, such as self-ordered nanoporous anodic aluminum oxide (AAO),[25] may contain isolated parallel cylindrical pores nearly uniform in diameter along their pore axes. However, the AAO pore arrays may exhibit a certain pore diameter dispersion in turn causing dispersion of the single-pore imbibition lengths $L_s$[26,27] and, consequently, imbibition front roughening.[23]

While even predictive understanding of imbibition on the single-pore level has remained scarce,[21,28] frequently porous matrices, such as controlled porous glass (CPG),[29] in which pore walls and pores form two interpenetrating networks, are imbibed.[30,31] Then, single-pore imbibition phenomena are superimposed by cooperative effects related to the presence of hydraulically coupled menisci at the imbibition front.[2,32-34] Cai and Yu state "…that most values of time exponent are less than 0.5. This may be explained that the LW equation may not be well suitable to depict the imbibition in a complex tri-dimensional porous medium. (…) The anomalous value of time exponent (<0.5) suggests that the imbibition speed becomes slower than that of pure considerations of capillary pressure and viscous drag as the average interface height increases."[35] If local variations of the pore diameters of the porous matrix exist at the imbibition front, menisci in different pores have different curvatures so that the Laplace pressure across these menisci varies. These pressure differences between hydraulically coupled menisci cause in turn cooperative changes of the imbibition front topography resulting in avalanche-like relaxations,[36-40] viscous fingering[36, 41] and imbibition front roughening.[26, 36, 41, 42] Thus, modified descriptions of imbibition dynamics considering tortuosity and geometry of the infiltrated pore system were suggested.[43]

To describe the spatiotemporal evolution of imbibition fronts during the invasion of porous matrices by wetting fluids, average positions of the imbibition fronts need to be determined and imbibition front widths need to be quantified in a statistically valid way. Imbibition fronts were, for example, inspected with the naked eye[44, 45] and by averaging methods, such as small-angle X-ray scattering,[21] dielectric spectroscopy,[46-48] neutron radiography,[26, 27] gravimetry[49-51] and opto-fluidic techniques.[20, 52-57] In particular, the reliable and reproducible quantification of imbibition front widths for systems characterized by imbibition front widths ranging from a few 100 nm to a few microns has remained challenging. Recently, we reported the statistical evaluation of 3D reconstructions of AAO membranes infiltrated with polystyrene (PS),[23] which were acquired by Zernike phase-contrast X-ray computed tomography[58-63] with submicron resolution. Here we compare the imbibition of PS into CPG, a pore model in which pores and pore walls form spongy interpenetrating networks, and into self-ordered AAO (Figure 1). Self-ordered AAO, a nearly ideal pore model with a tortuosity close to 1, contains arrays of straight parallel cylindrical pores oriented strictly normal to the membrane surface (Figure S1). The lengths of the percolation paths through a self-ordered AAO membrane along any of its pores correspond to the membrane thickness. The CPG and AAO membranes had both hydroxyl-terminated inorganic-oxidic pore walls and approximately similar mean pore diameters of ~380 nm. A pore diameter histogram of the self-ordered AAO membranes derived from a SEM image is shown in Figure 2a (reproduced from Supporting Figure S2 of our previous work[23]). A pore diameter histogram of the CPG membranes derived from mercury intrusion data is shown in Figure 2b. PS-infiltrated CPG and AAO membranes were investigated by Zernike phase-contrast X-ray computed tomography yielding volumetric reconstructions of extended sample volumes with spatial sub-micron resolution (Figure 1a, b) as well as, for comparison, by energy-dispersive X-ray (EDX) spectroscopy yielding 2D EDX maps (Figure 1c, d). In contrast to most averaging non-microscopic methods employed to monitor imbibition mentioned above, X-ray computed tomography enables probing imbibition-relevant structural features characterizing transient imbibition stages with single-pore resolution in the sub-µm range. On the other hand, much larger sample volumes than by classical microscopic techniques, such as electron microscopy, can be probed. Thus, structural features characterizing transient imbibition stages, such as imbibition lengths and imbibition front widths, can be imaged and evaluated with excellent validity even when they are characterized by submicron length scales. The only similar powerful method would be focus ion beam (FIB) tomography. However, FIB tomography requires the ablation of sample material between the imaging of consecutive slices, which may result in the emergence of artifacts. Moreover, in contrast to FIB tomography X-ray computed tomography allows the deployment of phase-contrast imaging, which is particularly suitable to image interfaces, such as imbibition fronts. We show that the obtained raw data can be analyzed by considering the pixel intensity dispersion $R_q$ parallel to the membrane surfaces as a function of the distance $D$ to the membrane surface. The real-space structures represented by the raw data are projected into 1D brightness dispersion profiles by a two-step (for 3D reconstructions) or a one-step (for 2D maps) procedure. Imbibition front positions and descriptors for imbibition front widths are then extracted from the 1D brightness dispersion profiles. Neither segmentation nor the identification of pore center coordinates are necessary. We show that the comparison with self-ordered AAO as a reference pore model having a tortuosity close to 1 yields information on the pore morphology of probed tortuous porous matrices, such as CPG, and insight into the mechanisms governing their imbibition.

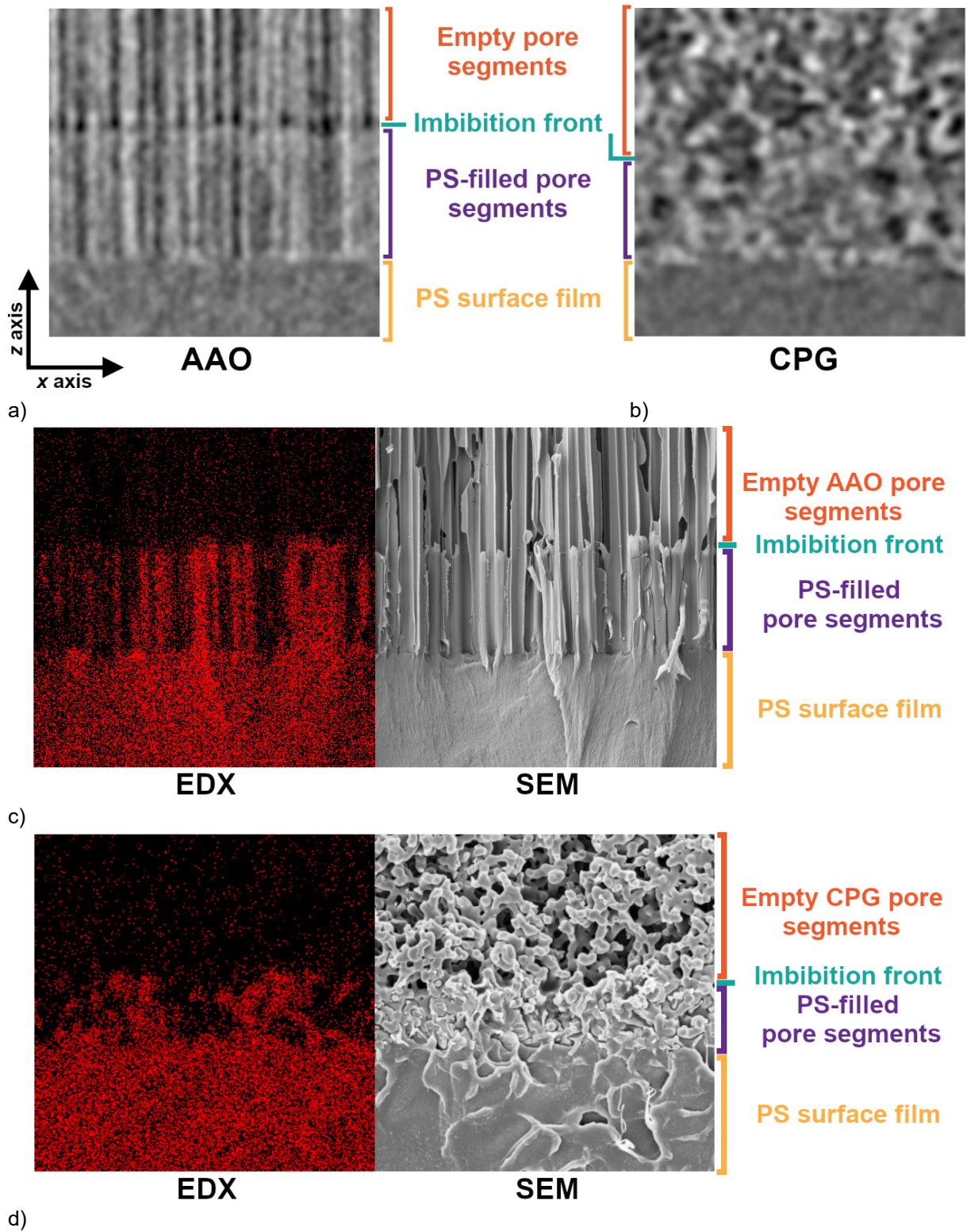

**Figure 1.** Cross sections normal to the membrane surface and parallel to the *xz* plane of a reference coordinate system (*x* axis parallel, *z* axis normal to the membrane surfaces) of a), c) AAO and b), d) CPG infiltrated with PS for $t_i$ = 3 min at 200°C. Empty parts of the membranes are at the top, membrane parts with PS-filled pore segments in the middle, and bulk PS surface films at the bottom. a), b) Parts of processed X-ray computed tomography *xz* slices. The edge lengths of the shown image fields along the *x* and *z* directions are 10.05 μm. c), d) 2D EDX maps extending 14.47 x 14.47 μm$^2$ showing the intensity of the carbon Kα peak and the corresponding SEM images.

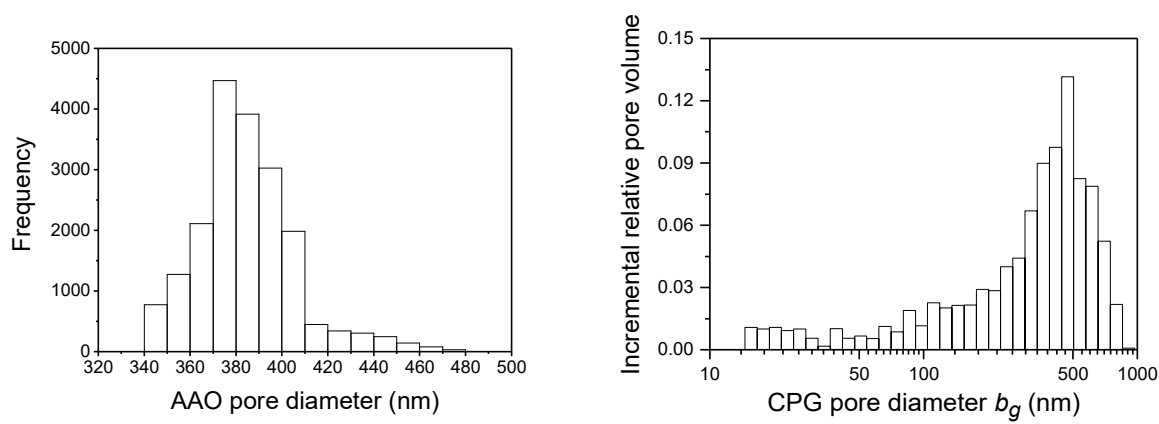

**Figure 2.** Pore diameter histograms. a) AAO pore diameter histogram (Figure S2 of Vazquez et al.[23]) obtained by the evaluation of a SEM image extending 2048 pixels x 1350 pixels showing an image field of 87.52 µm x 57.69 µm (23.4 pixels/µm) acquired with 1290fold magnification by using an InLens detector. The pore depth of the investigated AAO membrane amounted to 1 µm. b) CPG pore diameter histogram derived from a mercury intrusion measurement showing the portion of the relative pore volume occupied by pore segments with a diameter falling in a pore diameter increment centering about a pore diameter value $b_g$.

## Materials and Methods

### Materials

Monodisperse *sec*-butyl- and *H*-terminated PS (mass-average molecular weight $M_w$ = 239 kg/mol; number-average molecular weight $M_n$ = 233 kg/mol; PDI = 1.03) was supplied by Polymer Standards Service (Mainz, Germany). Toluene (anhydrous; 99.8 %) was purchased from Sigma Aldrich. Silicon wafers were purchased from Siegert Wafer GmbH. Self-ordered AAO was prepared following the procedures reported elsewhere.[23, 25]

### Preparation and characterization of CPG membranes

CPG membranes[29] with a thickness of 490 µm ± 10 µm extending 5 mm x 5 mm were produced by selective leaching applying a combined acid/alkaline treatment (1 M HCl, 90 °C, 2 h and 0.5 M NaOH, RT, 2 h) of phase-separated alkali borosilicate glass plates (composition: 62.5 wt.% $SiO_2$, 30.5 wt.% $B_2O_3$ and 7 wt.-% $Na_2O$; heat treatment 700 °C for 24 h), which results in the formation of spongy-continuous pore systems. The CPG membranes were characterized by mercury intrusion[64] (Figure 2b) in a pressure range from 1 to 1000 bar with a Pascal 440 porosimeter (ThermoScientific/Porotec). Prior to any measurement, the samples were degassed at 0.2 mbar and 23°C. To obtain a pore diameter histogram, the applied pressure is gradually increased, which causes gradual intrusion of mercury into the probed pore system. The relation between the applied pressure and the diameter of the pore segments filled at the applied pressure is inverse. The pore diameter is calculated by the Washburn equation using a mercury surface tension of 0.484 N/m and a mercury contact angle of 141.3°. The intruded incremental volume of mercury associated with an individual pressure step is the difference between the cumulative intrusion volumes after and prior to the application of the pressure step. The relative incremental pore volume ascribed to a pore diameter increment is the corresponding proportion of the total pore volume, which is the cumulative volume of mercury intruded into the probed pore structure at maximum pressure. Pore diameter histograms are obtained by plotting relative incremental intrusion volumes against the medium values of the corresponding pore diameter intervals. Porosities and specific pore volumes are calculated using the total pore volume. The CPG membranes used in this study have a median pore diameter ≈ 380 nm, 58 % porosity and a specific pore volume of 0.628 cm$^3$/g.

**Infiltration of PS**

The infiltration of AAO membranes attached to underlying 900 µm thick alumina substrates with PS was reported elsewhere.[23] In brief, AAO membranes attached to underlying aluminum substrates were heated to 200 °C in an argon atmosphere for 10 min. Then, unsupported 100 µm thick PS films were placed on the membrane surfaces for the desired infiltration times $t_i$ in an argon atmosphere. The infiltration was quenched by immersing the samples into ice-cooled bi-distilled water. For the infiltration of CPG membranes with PS the same procedure as for AAO[23] was applied, except that the freestanding PS films were produced with a smaller PS solution volume of 10 µL.

**X-ray computed tomography Imaging**

To detect the PS imbibition front, Zernike phase-contrast X-ray computed tomography imaging was performed using a Carl Zeiss Xradia Ultra 810 device operating at 5.4 keV (Cr X-ray source). The sample preparation for X-ray computed tomography was carried out as described elsewhere.[23] Samples were imaged with a field-of-view of 64 × 64 µm$^2$ without camera binning. A total of 901 projections with an exposure time of 60 s were collected over 180°. Image reconstruction was performed by a filtered back-projection algorithm using the software XMReconstructor integrated into the device, and tomograms obtained were exported as a stack of 16-bit TIFF images of approximately 1024 × 1024 pixels with an isometric pixel size of 64 nm. OriginPro 2020 and MATLAB R2021a were used for data analysis.

**EDX spectroscopy mappings of cross-sectional specimens**

Before imaging, the aluminum substrates of the PS-infiltrated AAO membranes were selectively etched with a solution containing 3.4 g CuCl$_2$•2H$_2$O and 100 mL HCl per 100 mL H$_2$O at 0°C. The freestanding PS-infiltrated AAO and CPG membranes were then cleaved perpendicularly to the membrane surface. The samples were sputter-coated with a platinum-iridium layer three times at 20 mA for 15 seconds in a K575X Emitech sputter coater. Scanning electron microscopy (SEM) images were obtained using a Zeiss *Auriga* microscope applying an acceleration voltage of 3 kV for AAO samples and 3.5 kV for CPG samples. Secondary electron secondary ion (SESI) and InLens detectors were used for image acquisition. Window integral EDX maps of the carbon Kα peak at 0.277 keV were obtained at a working distance of 5 mm using an EDX system Aztec (Oxford) equipped with an 80 mm$^2$-SDD-detector attached to the Zeiss *Auriga* SEM. Accelerating voltages of 3.0 kV and 3.5 kV were applied for the mappings of cross-sectional AAO and CPG samples. The 2D EDX maps extending 2048 pixels by 1532 pixels were imaged until at least 1 million counts were collected. The original EDX maps had an image depth of 32 bit. Pixel intensity dispersion profiles $R_{q,EDX}(D)$ along horizontal pixel rows in *x* direction of the external reference coordinate system parallel to the membrane surfaces, where *D* is the distance to the membrane surface along the *z*-direction of the reference coordinate system, were then extracted with the software Gwyddion.[65] The software Gwyddion read the EDX maps with an image depth of 8 bit corresponding to 256 shades of red from 0 to 255. For the visualization of the EDX maps, a pixel binning factor of 2 was applied to combine 4 adjacent pixels forming 2 x 2 square matrices into one new pixel using the Aztec software.

## Results and discussion

**Analysis of X-ray computed tomography data (method *Batch-CT-rms*)**

Because PS has a high glass transition temperature of ~100°C, transient imbibition stages can be frozen after selected imbibition times $t_i$ by thermal quenching. At room temperature, the glassy PS is vitrified and has solid-like properties so that the frozen transient imbibition stages can be conveniently characterized *ex situ*. We could thus obtain 3D reconstructions of AAO and CPG membranes infiltrated with PS for different imbibition times $t_i$ by X-ray computed tomography (method *Batch-CT-rms*). The 3D reconstructions consisting of 16-bit *xy* slices, which were approximately parallel to the membrane surfaces and the *xy* plane of the reference coordinate system (Figure S2), were statistically analyzed. At first, a suitable sub-volume was selected. Using the software ImageJ,[66] the original 16-bit *xy* slices were resliced so as to obtain *xz* slices and tilt-corrected using an ImageJ reference grid. As a result, the processed *xz* slices were oriented normal to the membrane surfaces, i.e., the membrane surfaces in the *xz* slices were horizontally oriented (parallel to the *x* direction) (Figure 1a, b). The *z*-axis was oriented normal to the membrane surface and, in the case of AAO, parallel to the AAO pore axes (Figure 1a). The selected volumes of interest of the *xz* slices contained portions of the PS surface film, PS-filled pore segments and empty pore segments, while visible large artifacts, such as the edges of the sample or large cracks (green boxes in Figures S2b and S2d) were excluded. Secondly, a MATLAB (R2021a, MathWorks Corporation) script was used to extract matrices from the processed *xz* slices, in which gray values were assigned all combinations of *x* and *z* values representing positions of pixels. The pixel gray values were then normalized to a scale ranging from 0 to 1. To extract information on the positions of the imbibition fronts and the membrane surfaces from *xz* slices, we tested different descriptors. The arithmetic mean value $I_{mean}$ of the pixel intensities along rows of pixels parallel to the membrane surfaces corresponding to the *x* direction of the reference coordinate system did not allow extracting meaningful information (Figure S3). This is intuitively understandable if one inspects *xz* slices with the naked eye (Figure 1a, b and Figure S2). Much better results were obtained using the pixel brightness dispersion $R_{q,s}$ along rows of pixels along the x-direction parallel to the membrane surface in the xz slices as descriptor. $R_{q,s}$ was calculated in the form of a root mean square roughness according to Equation 2:

$$R_{q,s} = \sqrt{\frac{1}{N}\sum_{j=1}^{N}(I_j - I_{mean})^2} \quad \text{(Equation 2)}$$

N is the number of pixels per row and $I_j$ the brightness value of pixel j ($1 \leq j \leq N$) in the considered row of pixels parallel to the x axis. Each row of pixels along the x-direction has a defined distance D along the z-axis to the membrane surface. Thus, for each xz slice a single-slice $R_{q,s}(D)$ profile was obtained, which represents the average pixel brightness dispersion along a pixel row in x-direction as a function of D. The single-slice $R_{q,s}(D)$ profiles are one-dimensional brightness dispersion tuples oriented normal to the membrane surfaces and parallel to the z-direction, which can be considered projections of the properties of the two-dimensional xz slices along the x-direction onto the z-axis of the reference coordinate system.

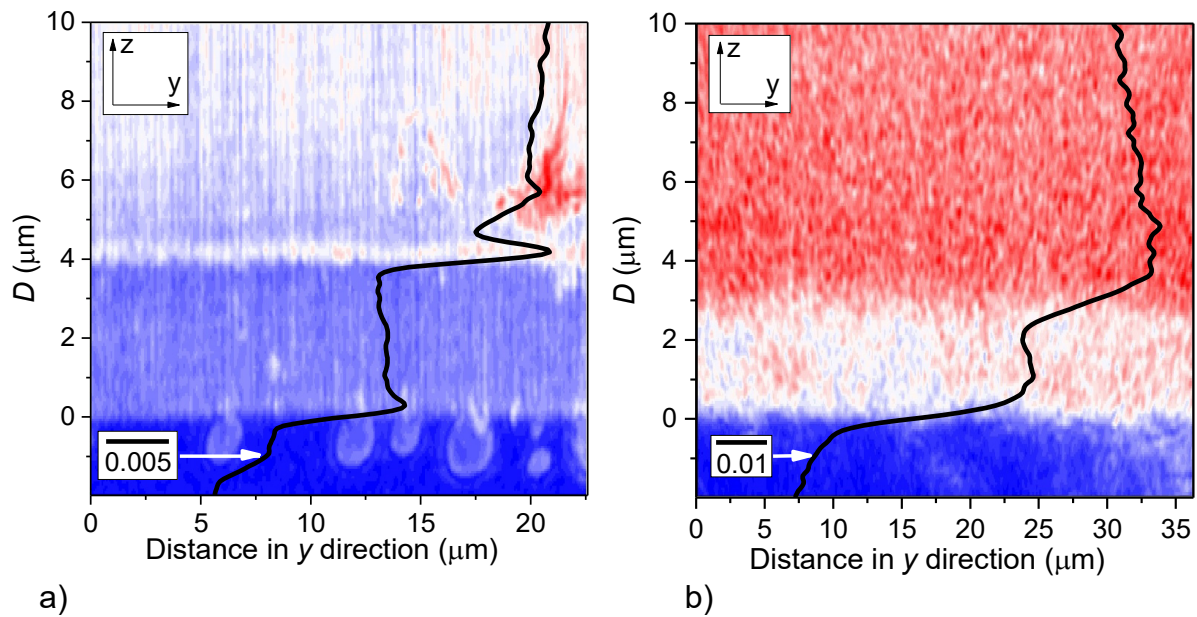

**Figure 3.** Statistical analysis of 3D reconstructions of volumes of interest obtained by X-ray computed tomography by means of yz color maps of a) an AAO membrane and b) a CPG membrane infiltrated with PS at 200°C for 3 minutes. In the yz color maps, each vertical pixel column represents a color-coded single-slice $R_{q,s}(D)$ profile of an xz slice through the 3D reconstructions. The distance D to the membrane surfaces is plotted along the vertical axis corresponding to the z-direction of the external reference coordinate system. Empty parts of the membranes are seen at the top, membrane parts with PS-filled pore segments in the middle, and the bulk PS surface films on the membrane surfaces at the bottom. The horizontal axis represents the y-direction of the reference coordinate system in the probed sample volume. The $R_{q,c}(D)$ profiles (black) are superimposed on the corresponding color map plots. The scale bars in the lower left-hand corners belong to the black $R_{q,c}(D)$ profiles.

Thirdly, all single-slice $R_{q,s}(D)$ profiles of the considered volume of interest were assembled into a two-dimensional array ("color map") lying in the *yz* plane of the reference coordinate system (Figures 3, S4 and S5). Each vertical pixel row parallel to the *z*-axis represents a single-slice $R_{q,s}(D)$ profile. The sequence in which the single-slice $R_{q,s}(D)$ profiles are arranged in the *yz* color maps corresponds to the real-space sequence of the corresponding *xz* slices. Therefore, the *yz* color maps are two-dimensional $R_{q,s}$ maps of the probed sample volume in the *yz* plane; for each *yz* position, the $R_{q,s}$ values representing the dispersion of the pixel brightness along the *x* direction are plotted in the *yz* plane.

Fourthly, color map $R_{q,c}(D)$ profiles (black curves in Figures 3, S4 and S5) were calculated for each pixel row in a color map parallel to the *y*-direction, that is, for each *D* value along the *z*-axis:

$$R_{q,c} = \frac{1}{N}\sum_{k=1}^{N} R_{q,s,k} \qquad \text{(Equation 3)}$$

In Equation 3, $N$ is the total number of considered *xz* slices corresponding to the total number of single slice $R_{q,s}(D)$ profiles assembled in the *yz* color maps. The index $k$ represents the number of the *xz* slice yielding the considered single-slice $R_{q,s}(D)$ profile in the stack of *xz* slices representing the real-space structure of the volume of interest. The $R_{q,c}(D)$ profiles were calculated using the original 16 bit gray-scale single-slice $R_{q,s}(D)$ profiles, i.e., the color coding of the single-slice $R_{q,s}(D)$ profiles was only used for the color map visualizations. The $R_{q,c}(D)$ profiles are one-dimensional projections of the properties of the *yz* color maps. Since the latter are in turn 2D projections of the properties of the volume of interest in the *x* direction parallel to the membrane surfaces, the $R_{q,c}(D)$ profiles are one-dimensional representations of the properties of the volume of interest in the direction normal to the membrane surface.

Figure 4a and b shows $R_{q,c}(D)$ profiles plotted against the distance *D* to the membrane surface. The volumes of interest covered by the AAO $R_{q,c}(D)$ profiles shown in Figure 4a contain ~2705 pores ($t_i$ = 3 min), ~7166 pores ($t_i$ = 10 min), ~3041 pores ($t_i$ = 20 min), ~2830 pores ($t_i$ = 30 min), ~3216 pores ($t_i$ = 70 min) and ~2961 pores ($t_i$ = 90 min). The pore numbers were estimated by assuming a perfect hexagonal pore lattice with a lattice constant of 500 nm. The volumes of interest covered by the CPG $R_{q,c}(D)$ profiles shown in Figure 4b amounted to ~32834 µm³ ($t_i$ = 3 min), 41940 µm³ ($t_i$ = 20 min), 53194 µm³ for ($t_i$ = 30 min) and 47108 µm³ ($t_i$ = 90 min). The three characteristic parts of PS-infiltrated AAO (Figure 3a, Figure S4) and CPG (Figure 3b, Figure S5) membranes are represented by three distinct segments of the $R_{q,c}(D)$ profiles (Figure 4). The areas representing the bulk PS surface films (bottoms in Figures 3, 4, S4 and S5) have the lowest pixel intensity dispersion, that is, the lowest $R_{q,c}(D)$ values. The surfaces of the AAO and CPG membranes separating the bulk PS surface film from PS-infiltrated membrane portions are marked by a stepwise increase (1) in the $R_{q,c}(D)$ profiles (cf. Figure 4). The higher $R_{q,c}(D)$ values in the membrane portions filled with PS are caused by the pixel intensity differences between pixels located in the PS-filled pore segments and pixels located in the oxidic pore walls (pixels intersected by the pore walls have intermediate pixel intensities[23]). The position of the imbibition front is marked by a second stepwise increase (2) in $R_{q,c}(D)$ (cf. Figure 4) because the difference in the pixel intensities between pixels located in the oxidic pore walls and in empty pore segments is even more pronounced than the difference in the pixel intensities between pixels located in the oxidic pore walls and in PS-filled pore segments. The stepwise increases (1) in $R_{q,c}(D)$ representing the membrane surfaces as well as the stepwise increases (2) in $R_{q,c}(D)$ profiles obtained from AAO samples representing the imbibition fronts in the AAO samples are superimposed by peaks. Only the stepwise increases (2) in $R_{q,c}(D)$ profiles obtained from CPG samples representing the imbibition fronts in the CPG samples have classical step-like shapes. Therefore, for AAO and CPG membranes different approaches were used to obtain the average imbibition lengths $L_a$ and measures of the imbibition front widths.

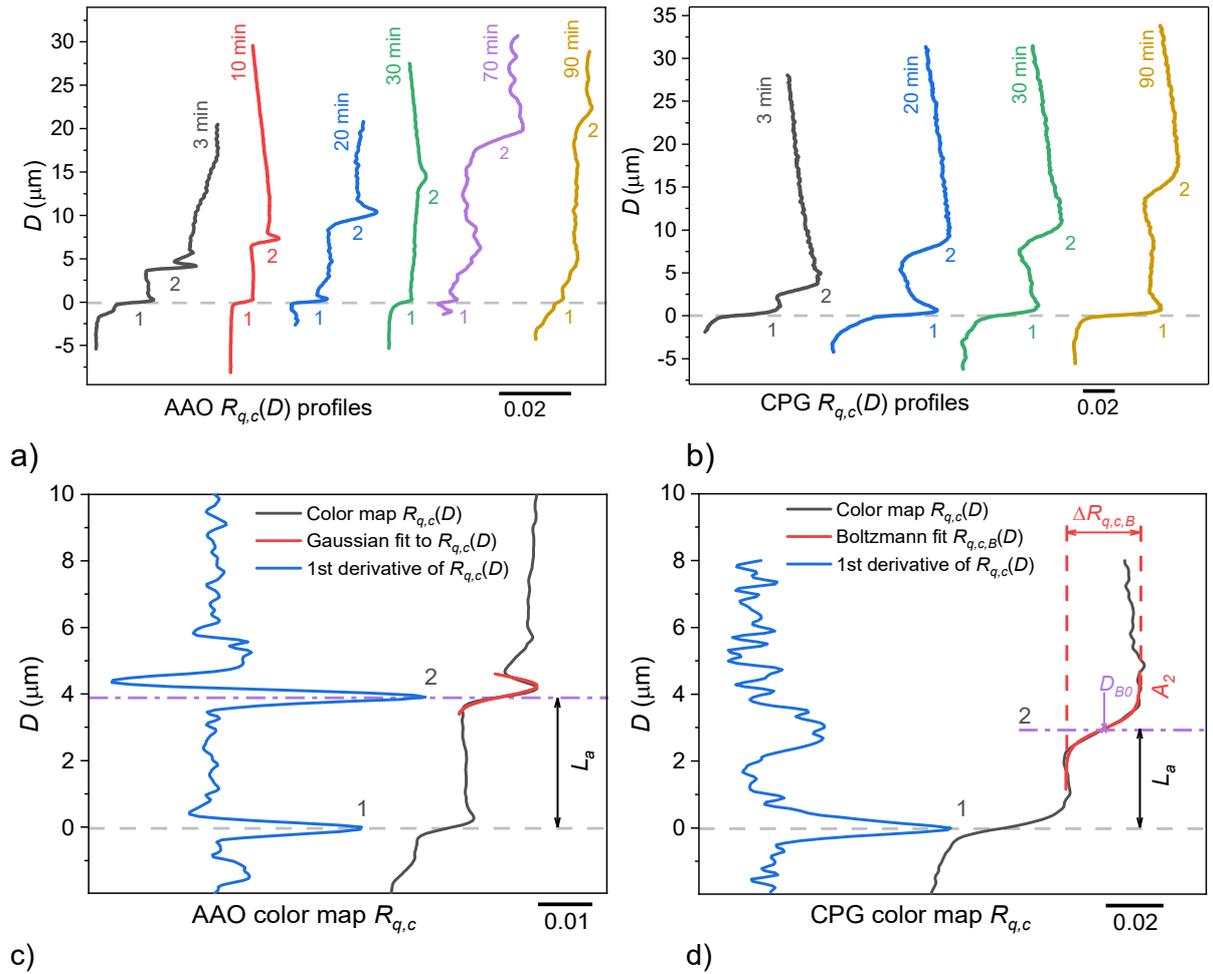

**Figure 4.** $R_{q,c}(D)$ profiles plotted against the distance $D$ to the membrane surface obtained for PS-infiltrated AAO and CPG membranes by evaluation of X-ray computed tomography data and their analysis to determine the average imbibition front position $L_a$ and to quantify the imbibition front width. The positions of the membrane surfaces at $D ≈ 0$ μm are indicated by dashed grey lines and the $R_{q,c}$ steps labelled (1). The $R_{q,c}$ steps labelled (2) indicate the average positions $L_a$ of the imbibition front. a) AAO $R_{q,c}(D)$ profiles. b) CPG $R_{q,c}(D)$ profiles. c) AAO $R_{q,c}(D)$ profile (black) for $t_i$ = 3 min, its first derivative $dR_{q,c}/dD$ (blue) and a Gaussian fit (red) to the $R_{q,c}(D)$ profile in the $D$ range where the imbibition front is located. d) CPG $R_{q,c}(D)$ profile (black) for $t_i$ = 3 min, its first derivative $dR_{q,c}/dD$ (blue) and a sigmoidal Boltzmann fit $R_{q,c,B}(D)$ (red) to the $R_{q,c}(D)$ profile in the $D$ range where the imbibition front is located. $D_{B0}$ is the $D$ value at the half height of $\Delta R_{q,c,B}$ step (2) of the sigmoidal Boltzmann fit indicating the position of the imbibition front (purple dash-dot line). $A_2$ is the maximum value of the sigmoidal Boltzmann fit $R_{q,c,B}$ representing the non-infiltrated part of the CPG membrane.

The average imbibition lengths $L_a$ for AAO samples were determined as illustrated in Figure 4c. $L_a$ was calculated as the difference between the $D$ values of the peak maximum positions of the 1st derivative $dR_{q,c}/dD$ indicating the presence of the imbibition front [peak (2), purple dashed-dot line] and the AAO surface at $D = 0$ μm [peak (1); grey dashed line]. As a descriptor of the imbibition front width, we determined the standard deviations $\sigma$ of Gaussian fits (red solid line in Figure 4c) to the peaks (2) of the $R_{q,c}(D)$ profiles.

The average imbibition lengths $L_a$ for CPG samples were obtained as illustrated in Figure 4d. The first derivatives $dR_{q,c}/dD$ of the $R_{q,c}(D)$ profiles were calculated. Then, the maxima of peaks (1) marking the CPG surface at $D = 0$ µm (grey dashed line) were determined. The steps (2) in the $R_{q,c}(D)$ profiles were fitted with a sigmoidal Boltzmann function:

$$R_{q,c,B}(D) = \frac{\Delta R_{q,c,B}}{1+\exp\left(\frac{D-D_{B0}}{k_B}\right)} + A_2 \quad \text{(Equation 4)}$$

$R_{q,c,B}(D)$ is the value of the sigmoidal Boltzmann function at a distance $D$ to the CPG surface (red solid line in Figure 4d). $\Delta R_{q,c,B}$ is the height of step (2) in the $R_{q,c,B}(D)$ curves and has a negative sign here. $A_2$ is the maximum value of $R_{q,c,B}$ representing non-infiltrated parts of the CPG membranes with empty pore segments, whereas $D_{B0}$ is the $D$ value at the half-heights of steps (2) in the $R_{q,c,B}(D)$ curves and represents the positions of the imbibition fronts. $L_a$ was determined as the difference between the $D_{B0}$ values of the sigmoidal Boltzmann fits to steps (2) and the maxima of peaks (1) of the first derivatives $dR_{q,c}/dD$ of the $R_{q,c}(D)$ profiles marking the CPG surfaces. The slope factor $k_B$ represents the steepness of step (2).

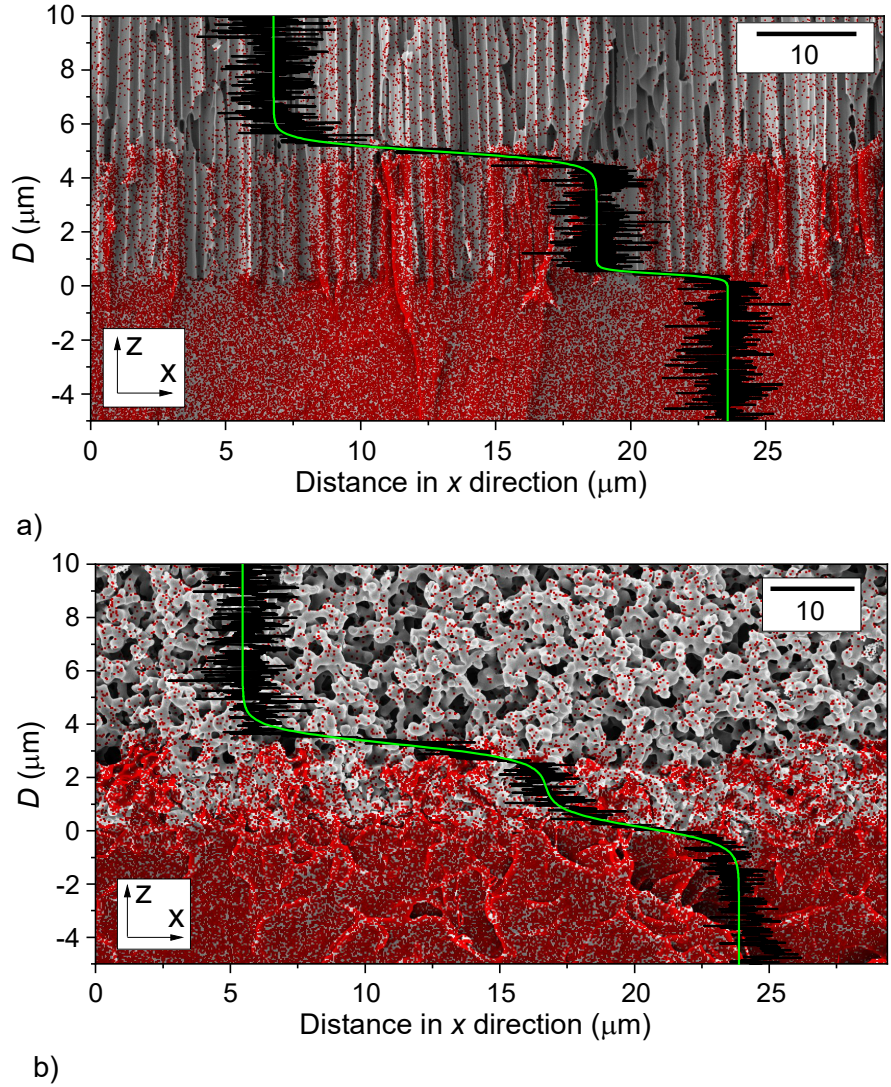

**Figure 5.** 2D EDX maps of the carbon Kα peak at 0.277 keV of cross-sections of a) an AAO membrane and b) a CPG membrane infiltrated with PS for 3 minutes at 200 °C. The EDX maps are superimposed on SEM images of the mapped areas. The distance $D$ to the membrane surface is plotted along the vertical axes corresponding to the z-axis of the external reference coordinate system. Empty parts of the membranes are seen at the top, membrane parts with PS-filled pore segments in the middle, and the bulk PS surface films on the membrane surfaces on the bottom. The noisy black curves superimposed on the EDX maps are $R_{q,EDX}(D)$ profiles of the dispersion $R_{q,EDX}$ of the pixel intensities on an 8-bit scale encompassing 256 shades of red. The $R_{q,EDX}(D)$ profiles represent the intensity of the carbon Kα peak along pixel rows in x-direction parallel to the membrane surfaces (in the case of AAO normal to the pores) at specific distances $D$ from the membrane surfaces. The smooth green curves are double-Boltzmann fits to the $R_{q,EDX}(D)$ curves calculated according to equation (6). The scale bars in the upper right-hand corners belong to the black $R_{q,EDX}(D)$ curves and their green double-Boltzmann fits. The length of the scale bars corresponds to $R_{q,EDX} = 10$.

**Analysis of 2D EDX maps (method *EDX-rms*)**

For comparison, we recorded 2D EDX maps of cross-sectional AAO (Figures 5a and S6) and CPG (Figures 5b and S7) specimens infiltrated with PS at 200°C for different imbibition times $t_i$ (method *EDX-rms*). The average position $L_a$ of the imbibition fronts, as well as measures of the imbibition front widths, were determined by statistical evaluation of 8-bit 2D EDX maps of the intensity of the carbon Kα peak indicating the distribution of PS, which predominantly consists of carbon (Figures S6 and S7). The 2D EDX maps were adjusted in such a way that the surfaces of the AAO and CPG membranes were parallel to the x-axis of the external reference coordinate system, whereas the distance D to the membrane surface was measured along the z-axis oriented perpendicularly to the membrane surfaces (and parallel to the AAO pores). The intensities of the carbon Kα peaks were encoded by 256 shades in a color space ranging from black (relative peak intensity zero) to red (relative peak intensity 1). As a measure of the pixel intensity dispersion, we calculated the average root mean square roughness $R_{q,EDX}$ of the pixel intensities along rows of pixels in the x-direction parallel to the surfaces of the AAO and CPG membranes using the software Gwyddion[65] according to:

$$R_{q,EDX} = \sqrt{\frac{1}{N}\sum_{j=1}^{N}(I_j - I_{mean})^2} \qquad \text{(Equation 5)}$$

N is the width of the evaluated 2D EDX maps in pixels, i.e., the number of pixels per row in the x direction parallel to the surfaces of the AAO and CPG membranes. The pixel intensity is denoted $I_j$, where j is the index of summation (1 ≤ j ≤ N), and $I_{mean}$ denotes the mean pixel intensity of all pixels in the considered row. Thus, for each pixel row at a certain distance D to the surfaces of the AAO and CPG membranes a $R_{q,EDX}$ (D) value was obtained (Figure 5). In the case of the 2D EDX maps the descriptors $I_{mean}$ and $R_{q,EDX}$ both allow the identification of the imbibition front and the membrane surface. However, $R_{q,EDX}$ profiles exhibit more prominent steps at the interface between PS-infiltrated and empty portions of the membranes. Therefore, $R_{q,EDX}$ is the more powerful descriptor (Figure S8). The $R_{q,EDX}$(D) profiles for all infiltration times $t_i$ exhibited two $R_{q,EDX}$ steps (Figure 6). The highest $R_{q,EDX}$ values were found in pixel rows located in the bulk PS films connected to the surfaces of the AAO and CPG membranes (bottoms of Figure 5a,b and 6a,b). While the entire probed volume consists of PS, local fluctuations in the carbon Kα peak intensity result in high $R_{q,EDX}$ values. The surfaces of the AAO and CPG membranes separating the bulk PS surface film from PS-infiltrated membrane portions are apparent by a stepwise decrease in $R_{q,EDX}$ (step (1) in Figure 6). The areas occupied by the pore walls are free of carbon. In these areas, no carbon Kα intensity was detected so that no local pixel intensity fluctuations occur. The imbibition front is indicated by a second stepwise decrease in $R_{q,EDX}$ (step (2) in Figure 6) since beyond the imbibition front only a small number of separated, scattered red pixels is present. Hence, the pixel rows predominantly consist of black pixels with little pixel intensity variations.

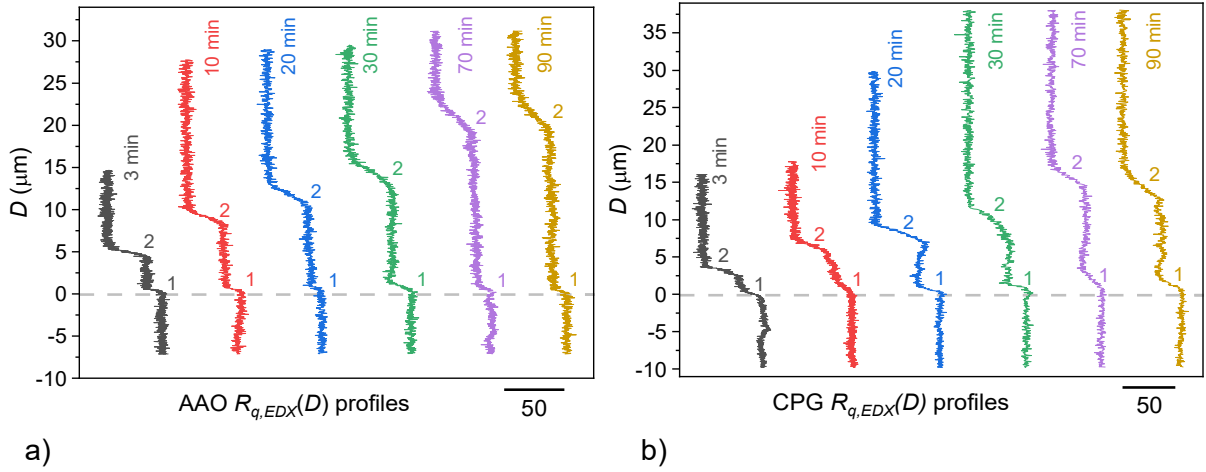

a)                                                                 b)

**Figure 6.** Statistical evaluation of pixel intensities representing the carbon Kα peak intensities in 2D EDX maps of cross-sectional specimens of a) AAO membranes and b) CPG membranes infiltrated with PS for different imbibition times $t_i$. The mean dispersion $R_{q,EDX}$ of pixel intensities along pixel rows in the x-direction parallel to the membrane surface is plotted against the distance $D$ to the membrane surface measured along the z-axis of the reference coordinate system. The $R_{q,EDX}$ steps labelled (1) at $D \approx 0$ µm (grey dashed line) mark the position of the membrane surface. The $R_{q,EDX}$ steps labelled (2) indicate the average position $L_a$ of the imbibition front. The AAO samples evaluated for panel a) extended 29.40 µm ($t_i$ = 3 min), 55.85 µm ($t_i$ = 10 min), 65.06 µm ($t_i$ = 20 min), 62.06 µm ($t_i$ = 30 min), 53.89 µm ($t_i$ = 70 min) and 54.37 µm ($t_i$ = 90 min) in the x direction parallel to the membrane surface. The CPG samples evaluated for panel b) extended 39.54 µm ($t_i$ = 3 min), 36.83 µm ($t_i$ = 10 min), 67.64 µm ($t_i$ = 20 min), 100.00 µm ($t_i$ = 30 min), 113.27 µm ($t_i$ = 70 min) and 100.39 µm ($t_i$ = 90 min) in the x-direction parallel to the membrane surface.

To calculate the average imbibition length $L_a$, we fitted a double sigmoidal Boltzmann function[67] to the $R_{q,EDX}(D)$ profiles (Figures 5 and 6):

$$R_{q,dB}(D) = R_{q,0} + A\left[\frac{p}{1+\exp\left(\frac{D-D_{EDX,1}}{k_{EDX,1}}\right)} + \frac{1-p}{1+\exp\left(\frac{D-D_{EDX,2}}{k_{EDX,2}}\right)}\right] \quad \text{(Equation 6)}$$

$R_{q,dB}(D)$ is the value of the double-Boltzmann fit function at a specific $D$ value (smooth green curves in Figure 5). $R_{q,0}$ is the $R_{q,dB}$ value in the $D$ range belonging to the bulk PS surface film covering the membrane surfaces. $R_{q,0} + A$ is the $R_{q,dB}$ value in the $D$ range belonging to the membrane parts with empty pore segments ahead of the imbibition front. $A$ is the sum of the heights of the two $R_{q,dB}$ steps (1) and (2) at the membrane surface and the imbibition front. $A \cdot p$ is the height of the $R_{q,dB}$ step (1) at the membrane surface and $A \cdot (1 - p)$ the height of $R_{q,dB}$ step (2) at the imbibition front. $D_{EDX,1}$ is the $D$ value at the half-height of $R_{q,dB}$ step (1), and $D_{EDX,2}$ is the $D$ value at the half-height of $R_{q,dB}$ step (2). $L_a$ was calculated as the difference between $D_{EDX,2}$ and $D_{EDX,1}$. The slope factors $k_{EDX,1}$ and $k_{EDX,2}$ represent the steepness of $R_{q,dB}$ steps (1) and (2) in the $R_{q,dB}(D)$ profiles.

**Mean imbibition front positions**

Self-ordered AAO and CPG are both inorganic oxidic pore models having hydroxyl-terminated pore walls with high specific surface energy so that the interactions between the pore walls and the polystyrene melt should be similar. The pore diameters of ~380 nm are one order of magnitude larger than the diameter of gyration of the PS used here.[23] The slopes $dL_a/dt_i^{1/2}$ of the linear fits to sets of $L_a(t_i^{1/2})$ data points (Figure 7a and Table 1), which correspond to the prefactor $\nu$ of the Lucas-Washburn law (equation 1) are measures of the imbibition velocities. The values obtained by methods *Batch-CT-rms* and *EDX-rms* for the AAO samples can directly be compared with the values obtained by methods *CT-rms* and *CT-mean* in our previous work.[23] The results obtained with these four methods are in good agreement. For the CPG samples, the results obtained by methods *Batch-CT-rms* and *EDX-rms* are likewise in line. A direct comparison of the slopes $dL_a/dt_i^{1/2}$ for AAO and CPG can be drawn by considering the values obtained with the same evaluation method. The slope $dL_a/dt_i^{1/2}$ obtained by method *Batch-CT-rms* for the AAO samples is 1.48 times higher than the slope $dL_a/dt_i^{1/2}$ obtained by method *Batch-CT-rms* for the CPG samples. In good agreement with this outcome, the slope $dL_a/dt_i^{1/2}$ obtained by method *EDX-rms* for the AAO samples is 1.53 times higher than the slope $dL_a/dt_i^{1/2}$ obtained by method *EDX-rms* for the CPG samples. The differences in the slopes $dL_a/dt_i^{1/2}$ can be rationalized by the different tortuosities of AAO and CPG. Unfortunately, there is no canonical definition of tortuosity. Frequently, the tortuosity is defined as the ratio of the lengths of a curvilinear percolation path and the linear distance between two points in a porous matrix.[68, 69] In the following, we will adhere to this definition. The average imbibition length $L_a$ is the linear distance $D$ between the membrane surface and the center of the imbibition front. For straight cylindrical pores, as in self-ordered AAO, the tortuosity is close to 1, and $L_a$ corresponds to the actual distance the invading fluid has to travel through the AAO pores to reach a distance $D$ from the membrane surface corresponding to $L_a$. The curvilinear percolation paths through the CPG pore systems, which are available to an invading fluid to reach a distance $D$ from the membrane surface corresponding to $L_a$, are larger than $L_a$. Hence, the phenomenological imbibition velocities and, therefore, the slopes $dL_a/dt_i^{1/2}$ are smaller for CPG than for AAO. The 1.5 times faster imbibition front movement in AAO relative to that in CPG with a porosity of 58 % used here suggests a CPG tortuosity on the order of 1.5. This value is indeed a reasonable estimate, as discussed below. At first glance, a tortuosity of 1.5 appears to be low for an isotropic pore model. However, it was shown that the nature of percolation paths through a pore model and, therefore, the tortuosity mainly depend on the porosity[70] (whereas the pore radius is typically not being considered a relevant parameter). CPGs with pore diameters of 10 - 20 nm but a porosity of 50 % – 60 % were reported to have tortuosities of 1.5 - 1.6, as determined by permeability measurements.[71, 72] Furthermore, Shelekhin et al. showed that the porosity and the tortuosity in porous glass membranes can be described by the percolation theory.[73] These authors defined the tortuosity as an empirical coefficient to describe the random orientation of the pores determined not only by the diffusion paths but also by the amount of throughout porosity and calculated the tortuosity based on the porosity (fraction of leachables). For low porosities (< 30 %) high tortuosities were obtained. In good agreement with the experimental value of 5.9, a tortuosity of 6.5 was calculated for VYCOR glass with a porosity of 31 %. The theoretical calculations also showed that an increase in the porosity of the membranes to 60 % results in a decrease of the tortuosity to 1.5.

Power laws like the Lucas-Washburn law consisting of a prefactor and a variable $t_i$ with an exponent $n$ having a value between 0 and 1 are generic descriptions of processes slowing down with time. The exponent $n$ of $t_i$ selected when $L_a$ is plotted against $t_i^n$ determines the slope $\mathrm{d}L_a/\mathrm{d}t_i^n = v$ of linear fits to sets of $L_a(t_i)$ data points. Pearson correlation coefficients $r$ of these linear fits are independent statistical measures of the linearity of the correlation between $L_a$ and $t_i^n$. The closer $r$ to 1 is, the higher is the linearity. In this way, it can be evaluated from a statistical point of view which exponent $n$ of $t_i$ yields the best description of the experimental results. This statistical evaluation reflects not only the physics underlying the probed imbibition process but also the quality of the measured data and the procedures for their evaluation. Therefore, we evaluated, which exponent $n$ of $t_i$ yields the best linear fit to sets of $L_a(t_i)$ data points by determining $r$ (Figure 7b).[74] It should, however, be noted that a precise determination of the optimum value of $n$ with an accuracy better than 0.1 is realistically not possible because the changes in $r$ may be rather minute. The analysis of 3D reconstructions of the CPG samples obtained by X-ray computed tomography using method *Batch-CT-rms* revealed that $r$ was closest to 1 for $n$ = 0.4, whereas the analysis of 3D reconstructions of the AAO samples obtained by X-ray computed tomography using method *Batch-CT-rms* revealed that $r$ was closest to 1 for $n$ = 0.45. Analysis of the EDX maps by method *EDX-rms* yielded the best $r$ values for $n$ values of 0.35 (AAO) and 0.3 (CPG). We interpret this outcome as follows. The exponents $n$ of $t_i$ yielding the best linear fits to the sets of $L_a(t_i)$ data points obtained by computed X-ray tomography (method *Batch-CT-rms*) are in reasonable agreement with the expectations emerging from physical considerations (see introduction). The exponent obtained for AAO with its separated cylindrical pores is close to 0.5, the exponents obtained for CPG are somewhat smaller. However, the exponents $n$ of $t_i$ obtained by statistical evaluation of results computed X-ray tomography (method *Batch-CT-rms*) and EDX mappings (method *EDX-rms*) differ to some extent. We assume that method *Batch-CT-rms* yields more valid results because much larger sample volumes are probed and preparation artifacts are prevented that may occur when infiltrated membranes are cleaved to prepare cross-sectional specimens for the EDX mappings.

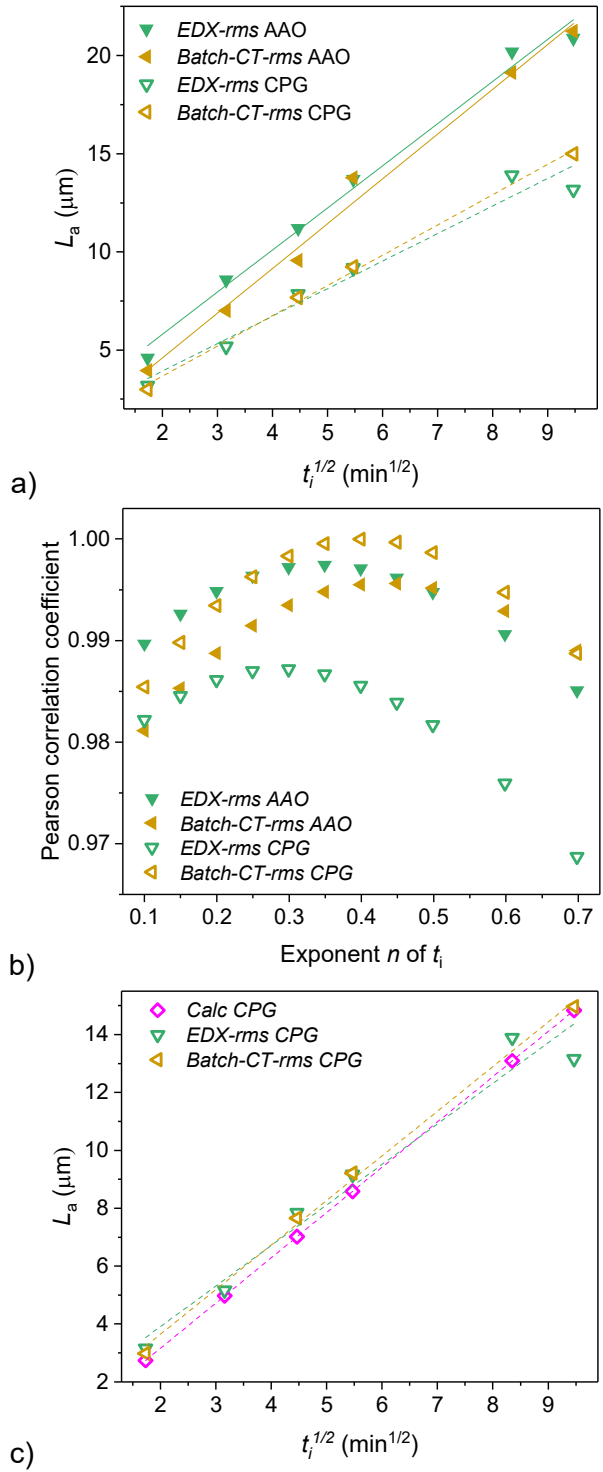

**Figure 7.** Comparison of the average imbibition lengths $L_a(t_i)$ obtained with methods *Batch-CT-rms* (ochre left-pointing triangles) and *EDX-rms* (green down-triangles) for AAO (solid symbols) and CPG (open symbols) membranes infiltrated with PS at 200°C. a) $L_a$ plotted against the square root of the imbibition time $t_i^{1/2}$. Solid lines are fits to sets of data points obtained with AAO membranes and dashed lines fits to sets of data points obtained with CPG membranes. b) Pearson correlation coefficients $r$ of linear fits to sets of $L_a(t_i^n)$ data points obtained by methods *EDX-rms* and *Batch-CT-rms* plotted against the exponent $n$ of $t_i$. c) Average imbibition lengths $L_a(t_i)$ in CPG membranes obtained by methods *Batch-CT-rms* and *EDX-rms* in comparison with results of equation (10) using a CPG pore size distribution obtained by a mercury intrusion measurement (pink diamonds). The dashed lines are linear fits to the sets of data points.

## CPG: role of dead layers

To consider "dead layers" of adsorbed PS molecules on the pore walls,[13-16] an effective pore radius $b_{eff}$ was introduced Yao et al.,[16] which is obtained by subtracting the thickness $\Delta b$ of the dead layer from the geometric pore radius $b_0$:

$$b_{eff} = b_0 - \Delta b \quad \text{(Equation 7)}$$

Thus, the prefactor $v$ of the Lucas-Washburn law for cylindrical pores as in AAO, which corresponds to the slopes $dL_a/dt_i^{1/2}$ of the linear fits to the experimentally determined sets of $L_a(t_i^{1/2})$ data points, can be expressed as:[23]

$$v = \sqrt{\frac{b_{eff}^4 \gamma \cos\theta}{2\eta_0 b_0^3}} \quad \text{(Equation 8)}$$

As discussed elsewhere,[23] the bulk viscosity $\eta_0$ of PS at the infiltration temperature of 200°C can be assumed to be $1.15 \cdot 10^4$ Pa·s. For the surface tension $\gamma$ of PS a value of 29 mJ/m² was used, and the contact angle $\theta$ of PS on the oxidic hydroxyl-terminated pore walls of the porous matrices used here was assumed to be 0°.[75] Hence, a comparison of equation (8) with the slopes $dL_a/dt_i^{1/2}$ of the linear fits to the experimentally determined sets of $L_a(t_i^{1/2})$ data points immediately yields $b_{eff}$ and, therefore, $\Delta b$. The evaluation of XRM data by an approach based on the identification of AAO pore centers reported previously by us thus yielded a $\Delta b$ value of ≈ 40 nm.[23]

For isotropic spongy-continuous pore networks, such as the pore networks of CPGs, the situation is more complex. It is reasonable to assume that only a certain proportion of the pores close to the imbibition front exhibits a pressure gradient caused by the presence of a mensiscus so that only this proportion of the pore contributes to the flow of the invading fluid. It was suggested to consider this aspect by dividing the flow rate by a hydrodynamic tortuosity $\tau_h$ (see, for example, Gruener et al.[76]). For isotropic spongy-continuous pore systems, the expression for the prefactor $v$ of the Lucas-Washburn law is, therefore, modified as follows:

$$v = \sqrt{\frac{b_{eff}^4}{\tau_h}} \cdot \sqrt{\frac{\gamma \cos\theta}{2\eta_0 b_0^3}} \quad \text{(Equation 9)}$$

Apart from $b_0$, the quantities in the second square root $\sqrt{\frac{\gamma \cos\theta}{2\eta_0 b_0^3}}$ are known from the literature. The mean pore radius $b_0$ of CPG was obtained from mercury intrusion data (Figure 2b). The prefactor $v$ of the Lucas-Washburn law was experimentally determined from the slopes $dL_a/dt_i^{1/2}$ of the linear fits to the sets of $L_a(t_i^{1/2})$ data points obtained by method *Batch-CT-rms*. While $b_{eff}$ and $\tau_h$ cannot be determined independently, equation (9) can easily be resolved for $\sqrt{b_{\text{eff}}^4/\tau_h}$ so that the value of this square root can be calculated. Thus, we obtained $\sqrt{b_{\text{eff}}^4/\tau_h} \approx 14699$ nm$^2$.

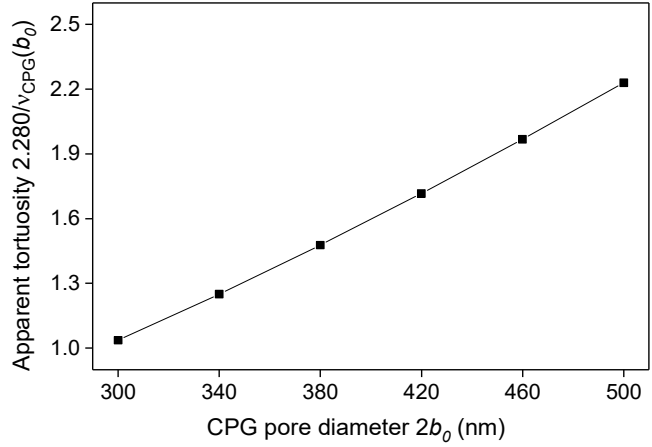

**Figure 8**. Apparent tortuosities 2.280/$v_{\text{CPG}}(b_0)$ apparent from the comparison of the imbibition of PS into CPG and into AAO as a reference pore model, which were calculated using equation (12), plotted against the CPG pore diameter $2b_0$.

Equation (9) relates the mean pore radius $b_0$ of the pore model used, which is typically accessible by independent experimental methods such as mercury intrusion, to the prefactor $v$ of the Lucas-Washburn law. As apparent from Figure 2b, mercury intrusion experiments do not only yield a mean pore radius $b_0$ for the CPG membranes but also the underlying pore radius frequency density. A relative frequency of occurrence $f_{p,g}$, here represented by a relative volume fraction, is assigned to pore segments with radii falling into a pore radius increment centering about a central pore radius value $b_g$. Thus, pseudo-single-pore imbibition lengths $L_s(b_g,t_i)$ can be calculated for each $b_g$ value and each imbibition time $t_i$ by replacing the mean pore diameter $b_0$ in equation (9) by $b_g$:

$$L_{s,g}(b_g, t_i) = \sqrt{\frac{b_{eff}^4}{\tau_h}} \cdot \sqrt{\frac{\gamma \cos\theta}{2\eta_0 b_g^3}} \cdot \sqrt{t_i} \qquad \text{(Equation 10)}$$

The overall number of considered $b_g$ values and, therefore, of $f_{p,g}(b_g)$ data points is $z$ so that $1 \leq g \leq z$. It should be noted that in equation (10) the constant value of 14699 nm$^2$ is used for the square root $\sqrt{b_{\text{eff}}^4/\tau_h}$. This is a simplification because in equation (10) $b_{eff}$ depends on $b_g$ rather than on $b_0$. Summation of the $L_s(b_g)$ values obtained for a specific imbibition time $t_i$ according to equation (10) weighed with the relative frequency $f_{p,g}$ displayed in the CPG pore diameter histogram shown in Figure 2b yields then the average imbibition length $L_a$ for the respective $t_i$ value:

$$L_a(t_i) = \sum_{g=1}^{z}(f_{p,g} \cdot L_{s,g}) = \sum_{g=1}^{z}\left(f_{p,g} \cdot \sqrt{\frac{b_{eff}^4}{\tau_h}} \cdot \sqrt{\frac{\gamma \cos\theta}{2\eta_0 b_g^3}} \cdot \sqrt{t_i}\right) \qquad \text{(Equation 11)}$$

The comparison of the sets of $L_a(t_i^{1/2})$ data points obtained with equation (11) and the sets of $L_a(t_i^{1/2})$ data points obtained by methods *EDX-rms* and *Batch-CT-rms* (Figure 7c) reveals that the results obtained by methods *EDX-rms* and *Batch-CT-rms* can be reproduced using CPG pore diameter frequency densities obtained by mercury intrusion. Equation (11) does not represent an independent approach to the determination of imbibition front positions but rather a cross-check for additional validation.

We also used equation (9) to estimate how variations in the assumed mean CPG pore diameter $2 \cdot b_0$ impact the tortuosity apparent from the comparison of PS infiltration into CPG and AAO as a reference pore model. To this end, we calculated the $L_a(t_i)$ profiles assuming values of 150 nm, 170 nm, 190 nm, 210 nm, 230 nm and 250 nm for the CPG pore radius $b_0$ according to

$$L_a(t_i) = \sqrt{\frac{b_{eff}^4}{\tau_h}} \cdot \sqrt{\frac{\gamma \cos\theta}{2\eta_0 b_0^3}} \cdot \sqrt{t_i} \qquad \text{(Equation 12)}$$

and determined for each $b_0$ value the slope $v_{CPG}(b_0)$ of linear fits to the corresponding sets of $L_a(t_i)$ data points corresponding to the Lucas-Washburn prefactors. To estimate the tortuosities $2.280/v(b_0)$ apparent from imbibition dynamics, we divided the slope $v_{AAO}$ = 2.280 μm/min$^{1/2}$ experimentally obtained for AAO by method *Batch-CT-rms* (Table 1) by the slopes $v_{CPG}(b_0)$ and plotted the resulting apparent tortuosities against the CPG mean pore diameter $2 \cdot b_0$. The tortuosity changes by 0.25 if the pore diameter $2b_0$ changes by 10 % (Figure 8).

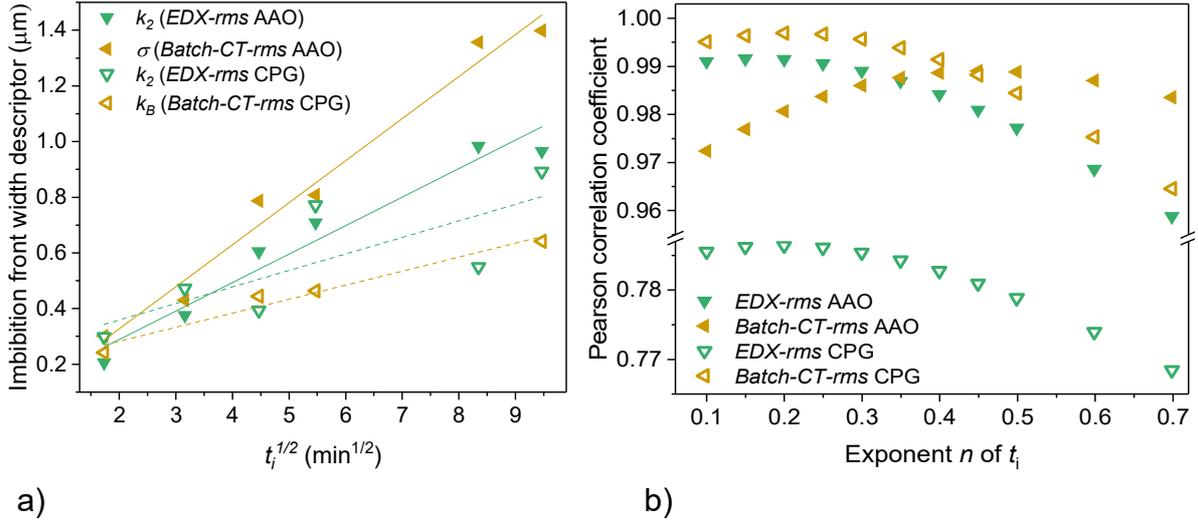

**Figure 9.** Comparison of the imbibition front width descriptors (*WD*) obtained by methods *Batch-CT-rms* (ochre left-pointing triangles) and *EDX-rms* (green down-triangles) for AAO (solid symbols) and CPG (open symbols) infiltrated with PS at 200°C. As *WDs* standard deviations $\sigma$ of Gaussian functions fitted to peaks (2) in $R_{q,c}(D)$ profiles (method *Batch-CT-rms* for AAO), slope factors $k_B$ of sigmoidal Boltzmann functions fitted to step (2) of $R_{q,c}(D)$ profiles (method *Batch-CT-rms* for CPG) and slope factors $k_2$ of double sigmoidal Boltzmann functions fitted to $R_{q,EDX}(D)$ profiles (method *EDX-rms* for AAO and CPG) were used. a) *WD* plotted against the square root of the imbibition time $t_i^{1/2}$. Solid lines are fits to sets of data points obtained with AAO membranes and dashed lines fits to sets of data points obtained with CPG membranes. b) Pearson correlation coefficients of linear fits to sets of $WD(t_i^n)$ data points obtained by methods *EDX-rms* and *Batch-CT-rms* plotted against the exponent $n$ of $t_i$.

**Imbibition front widths**

As discussed by Vazquez Luna et al.[23] (cf. section "Quantification of the Infiltration Front Width"), for imbibition processes and imbibition stages characterized by imbibition front widths from the sub-micron range to a few microns it is challenging, if not impossible, to reliably determine absolute imbibition front widths. Instead, statistical descriptors of imbibition front widths (width descriptors, *WD*) may be used, which can be extracted from experimental raw data by means of defined algorithms. The $R_{q,c}(D)$ profiles obtained by method *Batch-CT-rms* illustrate that specifically adapted descriptors for the imbibition front widths are required. In the case of the $R_{q,c}(D)$ profiles obtained by method *Batch-CT-rms* for the AAO samples the position of the imbibition front is marked by peaks (2) rather than by steps (Figure 4a). Therefore, in this case standard deviations $\sigma$ of Gaussian fits to the $R_{q,c}(D)$ profiles in a $D$ range around $L_a$ were used as *WD* (Figure 4c). However, step (2) in the $R_{q,c}(D)$ profiles obtained for CPG samples (Figure 4b) was fitted by a sigmoidal Boltzmann function (Figure 4d). In this case, the slope factor $k_B$ is a suitable *WD*. If $k_B$ approaches infinity, the slope of step (2) approaches zero, i.e., step (2) vanishes. If $k_B$ approaches zero, the slope of step (2) approaches infinity, i.e., step (2) is then vertical. The smaller $k_B$ is, the steeper is step (2) and the narrower is the imbibition front. As $k_B$ increases, the steepness of step (2) decreases and the imbibition front width increases. In the case of method *EDX-rms*, the imbibition front in the AAO as well as the CPG membranes is indicated by step (2) in the sigmoidal double-Boltzmann functions fitted to the $R_{q,EDX}(D)$ profiles obtained by evaluation of EDX maps (Figure 5). In this case, the slope factor $k_{EDX,2}$ of step (2), which has the same properties as slope factor $k_B$, is used descriptor of the imbibition front width. We used the slopes $dWD/dt_i^{1/2}$ of linear fits to sets of $WD(t_i^{1/2})$ data points to comparatively quantify the spatiotemporal evolution of the imbibition front width (Figure 9a). However, only $WD(t_i^{1/2})$ data points obtained with slope factors $k_B$ or $k_{EDX,2}$ as $WD$ can be compared directly. The slope $dk_B/dt_i^{1/2}$ to sets of $k_B(t_i^{1/2})$ data points obtained for CPG samples by method *Batch-CT-rms* amounts to 0.50 µm s$^{-1/2}$, while method *EDX-rms* yielded for the CPG samples $dk_{EDX,2}/dt_i^{1/2}$ = 0.59 µm s$^{-1/2}$. Furthermore, the $dk_{EDX,2}/dt_i^{1/2}$ value obtained for the CPG samples amounted to 58 % of the value obtained for the AAO samples (as compared to 65–68 % for the $dL_a/dt_i^{1/2}$ values).

We evaluated which exponent $n$ of $t_i$ yields the best linear fit to sets of $WD(t_i^n)$ data points (Figure 9b). For this purpose, we determined the Pearson correlation coefficient $r$ of linear fits obtained for different values of the exponent $n$ of $t_i$. An $n$ value of 0.20 yielded the best linear fits to sets of data points obtained with methods *Batch-CT-rms* and *EDX-rms* for the CPG samples. However, due to the minute $r$ variations for $n$ values smaller than 0.30 it is difficult to unambiguously determine optimum $n$ values. Moreover, the linear fits to the data points resulting from method *EDX-rms* have low $r$ values smaller than 0.79. For the AAO samples, methods *Batch-CT-rms* and *EDX-rms* yielded inconclusive results. Method *EDX-rms* yielded an optimum $n$ value of 0.15, albeit also here the differences between the $r$ values in the $n$ range below 0.3 are small. In contrast, method *Batch-CT-rms* yielded an optimum $n$ value of 0.45 (with little differences in $r$ in the range $0.4 \leq n \leq 0.5$), which is in good agreement with the values obtained previously with methods *CT-rms* and *CT-mean*.[23]

**Table 1.** Pearson correlation coefficients $r$ and slopes $\nu = dL_a/dt_i^{1/2}$ of the linear fits $L_a(t_i^{1/2})$ in Figure 7a as well as Pearson correlation coefficients $r$ and slopes $dWD/dt_i^{1/2}$ of the linear fits to the used imbibition front width descriptors ($WD$) plotted against $t_i^{1/2}$ data sets in Figure 9a. The values for methods *CT-rms* and *CT-mean* were taken from Vazquez Luna et al.[23] As imbibition front width descriptors, σ (method *Batch-CT-rms* for AAO), $k_B$ (method *Batch-CT-rms* for CPG), $k_2$ (method *EDX-rms* for AAO and CPG) and the descriptors described by Vazquez Luna et al.[23] (methods *CT-rms* and *CT-mean*) were used.

| Porous Material | Method | $r$ value $L_a(t_i^{1/2})$ | $\nu$ [µm/min$^{1/2}$] | $r$ value $WD(t_i^{1/2})$ | $dWD/dt_i^{1/2}$ [µm/min$^{1/2}$] |
|---|---|---|---|---|---|
| AAO | Batch-CT-rms | 0.995 | 2.280 | 0.989 | 0.151 |
| AAO | EDX-rms | 0.994 | 2.148 | 0.977 | 0.102 |
| AAO | CT-rms | 0.996 | 2.380 | 0.929 | 0.146 |
| AAO | CT-mean | 0.991 | 2.372 | 0.994 | 0.205 |
| CPG | Batch-CT-rms | 0.999 | 1.542 | 0.984 | 0.050 |
| CPG | EDX-rms | 0.981 | 1.402 | 0.778 | 0.059 |

For method *Batch-CT-rms* sample volumes of at least 30 000 µm$^3$ were probed – in contrast to method *EDX-rms* yielding only 2D maps. Hence, we assume that the validity of method *Batch-CT-rms* is better, which is in particular of importance when imbibition front widths in the micron and submicron range are evaluated. We thus conclude that the imbibition front width in AAO increases faster with $t_i$ than in CPG (Figure 9a). Moreover, the powers $n$ of $t_i$ in the power laws describing the scaling of the imbibition front width descriptors WD with $t_i$ determined by method *Batch-CT-rms*, which are associated with the best Pearson correlation coefficients, amount to 0.45 for AAO and 0.20 for CPG. This outcome is consistent with the results of previous studies on imbibition front broadening. In arrays of independent cylindrical pores exhibiting a certain pore diameter dispersion within the pore array, whereas individual pores are uniform in diameter, as it is the case for AAO, the menisci located in the separated pores move independently. If the movements of the menisci follow a Lucas-Washburn dynamics, the imbibition front roughening scales with the square root of the imbibition time $t_i$.[26, 27] Moreover, imaging of spontaneous imbibition of water into nanoporous Vycor glass with pore diameters below 10 nm using reflected light and neutrons revealed that imbibition front roughening scaled with the square root of $t_i$.[26] This outcome was traced to independent menisci movements in elongated pore segments. However, a simulation of imbibition front broadening based on a pore-network model shows a transition of kinetic roughening at a critical pore aspect ratio in interconnected pore networks.[26] As a result, the imbibition front roughening scales with $t_i$ raised to the power of an exponent smaller than 0.5 in spongy-continuous pore networks without neck-like pore segments, such as in the CPGs used here. In this case, the menisci do not move independently along the pores. Instead, an effective surface tension counteracts the broadening mechanisms.[26, 30] This outcome is in turn compatible with the visual inspection of SEM images and the relatively low tortuosity of 1.5 assumed for the CPG used here.

## Conclusion

We have comparatively studied the imbibition of polystyrene into AAO containing isolated straight cylindrical pores and CPG containing spongy-continuous pore systems. Both pore models have hydroxyl-terminated oxidic pore walls and pore diameters of ~380 nm, which exceed the dimensions of the infiltrated macromolecules by one order of magnitude. To statistically analyze 3D reconstructions of imbibed membranes obtained by Zernike phase-contrast X-ray computed tomography and, for comparison, 2D EDX maps, the pixel intensity dispersion parallel to the membrane surfaces was evaluated. It appears that the pixel brightness dispersion is a more powerful and more universal descriptor of the image properties than the mean pixel intensity when it comes to the identification of the positions of imbibition fronts and membrane surfaces. The 3D reconstructions and the 2D EDX maps were condensed into 1D profiles normal to the membrane surfaces representing the pixel intensity dispersion as a function of the distance to the membrane surface. To process the 3D reconstructions in this way, a two-step algorithm was developed comprising the successive calculation of pixel intensity dispersion profiles along two orthogonal directions parallel to the membrane surface. The approach reported here does neither involve segmentation procedures nor require the determination of pore coordinates. The statistical analysis of the 1D brightness dispersion profiles yielded consistent results regarding the imbibition front positions. The prefactor $v$ of the Lucas-Washburn law relating the imbibition front position to the imbibition time found for the AAO samples is 1.5 times larger than that found for the CPG samples. This outcome is in good agreement with the tortuosity of CPG reported earlier. Thus, imbibition retardation with respect to AAO reference samples having a tortuosity close to 1 may quantify the tortuosity of a probed pore system on condition that surface interactions are comparable. Using the prefactor $v$ experimentally determined from X-ray computed tomography data and CPG pore diameter frequency densities obtained by mercury intrusion, the average imbibition front positions in CPG could be reproduced by the Lucas-Washburn law. Analysis of the 1D brightness dispersion profiles also yields statistical descriptors of the imbibition front widths, the dependence of which on the imbibition times reveals to what extent individual or cooperative meniscus movements govern imbibition. X-ray computed tomography data indicate that in contrast to Vycor glass with sub-10-nm pores, meniscus movements in the CPGs used here are governed by cooperative effects. This outcome in turn indicates that neck-like cylindrical pore segments do not significantly impact the spatiotemporal imbibition front evolution in the CPGs tested here. The results presented here may help achieve predictive understanding of imbibition into real-life materials with spongy-continuous pore systems relevant to applications including printing, adhesive bonding, and the fabrication of hybrid materials.

ASSOCIATED CONTENT

**Supporting Information.** Large-field cross-sectional SEM image of self-ordered AAO, raw slices of 3D reconstructions obtained by X-ray computed tomography of infiltrated AAO and CPG membranes, evaluation of X-ray computed microscopy raw data and EDX maps using the mean pixel intensity as descriptor, *yz* color maps of infiltrated AAO and CPG membranes, 2D EDX maps of infiltrated AAO and CPG membranes. This material is available free of charge via the Internet at http://pubs.acs.org."

AUTHOR INFORMATION
**Corresponding Author**
* Martin Steinhart, Institut für Chemie neuer Materialien und CellNanOs, Universität Osnabrück, Barbarastr. 7, 49076 Osnabrück, Germany; e-mail: martin.steinhart@uos.de

**Present Address**
† Fraunhofer Institute for Microstructure of Materials and Systems IMWS, Walter-Hülse-Straße 1, 06120 Halle (Saale), Germany

## Author Contributions

The manuscript was written through contributions of all authors. All authors have given approval to the final version of the manuscript.

## Funding Sources

The authors thank the German Research Foundation for funding (PAK 949 "Nanostructured Glasses and Ceramics", project number 383411810, HU 850/9-1, STE 1127/19-1, WE 4051/22-1 and "High-throughput, Chemical X-ray Microstructure Screening Center for Functional Glasses and Glass Ceramics", project number 316987262, WE 4051/21-1).

## Notes

The authors declare no competing financial interest.

## ACKNOWLEDGMENT

The authors thank Dr. Cristine S. de Oliveira, Claudia Hess and Christine Schulz-Kölbel for technical support.